\documentclass{article}
\usepackage{amssymb}
\usepackage{amsthm}
\usepackage[dvips]{graphicx}

\def\pn{\mathrel{{\mathop+}_n}}
\def\mn{\mathrel{{\mathop-}_n}}
\def\tn{\mathrel{{\mathop\cdot}_n}}
\def\tnd{\mathrel{{\mathop\cdot}_{n/d_i}}}
\newcommand{\floor}[1]{\left\lfloor#1\right\rfloor}
\newcommand{\ceil}[1]{\left\lceil#1\right\rceil}
\newtheorem{fact}{Fact}
\newtheorem{lemma}[fact]{Lemma}
\newtheorem{theorem}[fact]{Theorem}
\newtheorem{corollary}[fact]{Corollary}

\begin{document}

\title{Minimal Chordal Sense of Direction and Circulant Graphs}

\author{Rodrigo~S.~C.~Le\~ao\thanks{Corresponding author
({\tt rleao@cos.ufrj.br}).}\\
Valmir~C.~Barbosa\\
\\
Universidade Federal do Rio de Janeiro\\
Programa de Engenharia de Sistemas e Computa\c c\~ao, COPPE\\
Caixa Postal 68511\\
21941-972 Rio de Janeiro - RJ, Brazil}

\date{March 3, 2005}

\maketitle

\begin{abstract}
A sense of direction is an edge labeling on graphs that
follows a globally consistent scheme and is known to considerably
reduce the complexity of several distributed problems. In this paper, we study aparticular instance of sense of direction, called a chordal sense of direction
(CSD). In special, we identify the class of $k$-regular graphs that admit a CSD
with exactly $k$ labels (a minimal CSD). We prove that connected graphs in this class
are Hamiltonian and that the class is equivalent to that of circulant graphs,
presenting an efficient (polynomial-time) way of recognizing it when the graphs'degree $k$ is fixed.

\end{abstract}

\section{Introduction}

In this paper we model a distributed system as an undirected graph $G$
on $n$ vertices having no multiple edges or self-loops. Every edge of $G$ is
assumed to have two labels, each corresponding to one of its end vertices. For
terminology or notation on graph theory not defined here we refer the reader to
\cite{bondy}. 

In \cite{santoro}, a property of this edge labeling was introduced which can
considerably reduce the complexity of many problems in distributed computing
\cite{flocchini7}. This property refers to the ability of a vertex to
distinguish among its incident edges according to a globally consistent scheme
and is formally described in \cite{flocchini1}. An edge labeling for which the
property holds is called a \emph{sense of direction} and is necessarily such
that all the edge labels corresponding to a same vertex are distinct (the edge
labeling is then what is called a \emph{local orientation}). We say that a sense
of direction is \emph{symmetric} if it is possible to derive the
label corresponding to one end vertex of an edge from the label corresponding to
the other. We say that it is \emph{minimal} if it requires exactly
$\Delta(G)$ distinct labels, where $\Delta(G)$ is the maximum degree in $G$.
For a survey on sense of direction, we refer the reader to \cite{flocchini8}.

\begin{figure}[t]
\centering
\begin{tabular}{c@{\hspace{1.5cm}}c}
\includegraphics[scale=0.75]{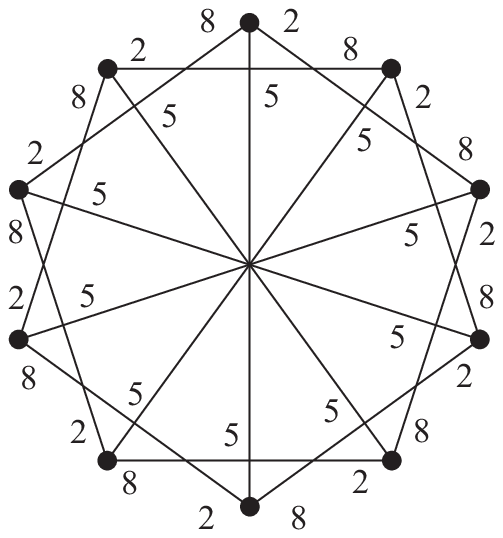}&
\includegraphics[scale=0.75]{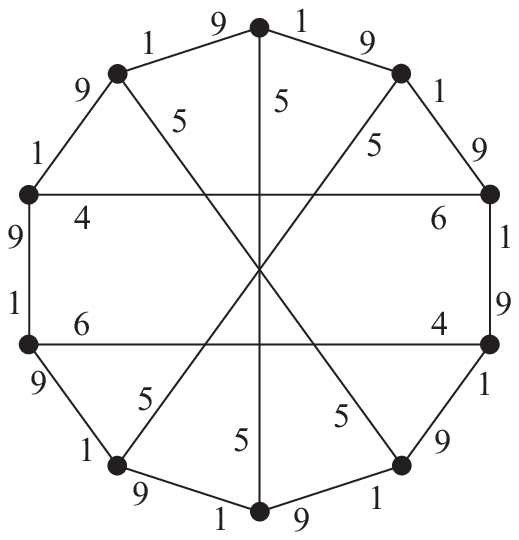}\\
{\small (a)}&{\small (b)}
\end{tabular}
\caption{A graph with an edge labeling that is an MCSD (a) and another graph
with an edge labeling that is a CSD but not an MCSD (b). Vertices are ordered
clockwise.}
\label{fig_example}
\end{figure}

A particular instance of symmetric sense of direction, called a \emph{chordal
sense of direction} (CSD), can be constructed on any graph by fixing an arbitrary
cyclic ordering of the vertices and, for each edge $uv$, selecting the
difference (modulo $n$) from the rank of $u$ in the ordering to that of $v$ as
the label of $uv$ that corresponds to $u$ (likewise, the label that corresponds
to $v$ is the rank difference from $v$ to $u$). In Figure~\ref{fig_example}(a),
an example is given of a minimal chordal sense of direction (MCSD). It is
relatively easy to see that there exist graphs that do not admit an MCSD, as for
instance the one in Figure~\ref{fig_example}(b). 

Given a finite group $A$ and a set of generators $S\subseteq A$, a \emph{Cayley graph}
is a graph $H$ whose vertices are the elements of the group ($V(H)=A$)
and whose edges correspond to the action of the generators
($uv \in E(H) \iff \exists s \in S: v=s*u$, where $*$ is the operation defined
for $A$). We assume that the set of generators is closed under inversion, so $H$
is an undirected graph. An edge labeling of $H$ assigning two labels to each edge
in such a way that each of an edge's labels corresponds to one of its end vertices
is called a \emph{Cayley labeling} if, for edge $uv$, the
label that corresponds to vertex $u$ is $s$ such that $v=s*u$. 

In \cite{flocchini6}, it was shown that a regular graph's edge labeling 
is a symmetric sense of direction if and only if the graph is a Cayley 
graph and the labeling is a Cayley labeling. This result was later extended to 
directed graphs in \cite{boldi3}, where the problem of recognizing labeled 
Cayley graphs was also demonstrated to be solvable in parallel polynomial time. 
This latter result uses the same $O(n^{14.256})$-time algorithm of 
\cite{boldi1}, where the problem of deciding whether a given labeling is a sense
of direction of a given graph was solved.

A \emph{circulant graph} (also known as a \emph{chordal ring}) is a Cayley graph
over $\mathbb{Z}_n$, the cyclic group of order $n$ under the addition operation.
The relevance of circulant graphs is due to their connectivity properties (small
diameter, high symmetry, etc.), which render them excellent topologies for
network interconnection, VLSI, and distributed systems \cite{bermond}.
The problem of recognizing circulant graphs is still challenging: results
are known only for very specific instances, like the cases of $n$ prime
\cite{muzychuk2}, geometric circulant graphs \cite{muzychuk6}, and recursive
circulant graphs \cite{fertin}.

In this paper, we identify and analyze the regular graphs that admit an MCSD. 
We describe their structure, show that they are all Hamiltonian if connected, and moreover
demonstrate an equivalence between certain distinct labelings.
We also show that the class of regular graphs that admit an 
MCSD and the class of circulant graphs are equivalent to each other. A
straightforward consequence of our analysis is that the problem of recognizing
circulant graphs can be polynomially solved when the graphs' degree is fixed.

Throughout the text, the operators $\pn$, $\mn$, and $\tn$ represent, 
respectively, the modulo-$n$ operations of addition, subtraction, and
multiplication.

\section{MCSD's of Regular Graphs}

Let $G$ be a $k$-regular graph that admits an MCSD, $\lambda$ a labeling that is 
an MCSD of $G$, and $\Gamma\subseteq\{1,\ldots,n-1\}$ the set of labels used by 
$\lambda$. Since $\lambda$ is minimal, we may write $\Gamma=\{\gamma_1,\ldots,
\gamma_k\}$ and assume, further, that $\gamma_1<\cdots<\gamma_k$.
We denote by $\lambda_u(uv)$ the label of edge $uv$ that corresponds
to vertex $u$. We also write $\lambda(uv)=\{\lambda_u(uv),\lambda_v(uv)\}=
\{\gamma_i,\gamma_j\}$ with $1\leq i,j\leq k$. It is easy to see 
that for any CSD there exists a symmetry function $\psi$ such that $\psi(\lambda_u(uv))=
\lambda_v(uv)$, given by $\psi(\gamma_i)=n-\gamma_i$ for $\gamma_i\in\Gamma$. 
We start by highlighting an important property of $\lambda$.

\begin{lemma}
\label{labeling}
If $k$ is even, then the edges of $G$ are labeled by $\lambda$ with the label 
pairs $\{\gamma_1,n-\gamma_1\},\ldots,\{\gamma_{k/2},
n-\gamma_{k/2}\}$.
If $k$ is odd, then a further label pair is
$\{\gamma_{\ceil{k/2}},n-\gamma_{\ceil{k/2}}\}=\{n/2,n/2\}$.
\end{lemma}
\begin{proof}
The $k/2$ label pairs for the case of $k$ even follow directly from 
the definition of $\psi$ and from the fact that $|\Gamma|=k$. If $k$ is odd, 
then the label $\gamma_{\ceil{k/2}}=n/2$ (necessarily an integer, since $k$ 
odd implies $n$ even when $G$ is $k$-regular) remains unused by any of those 
pairs, so a further label pair is $\{n/2,n/2\}$. 
\end{proof}

By Lemma~\ref{labeling}, we can always refer to $\lambda$ by simply giving the 
$\ceil{k/2}$ labels that are no greater than $n/2$. 
Having established this property of $\lambda$, we now set out to describe more 
about the structure of graphs that admit an MCSD. In what follows, we say that a 
graph $H$ \emph{decomposes} into the two subgraphs $A$ and $B$ when $V(A) \cup V(B)=
V(H)$, $E(A)\cup E(B)=E(H)$, and $E(A)\cap E(B)=\emptyset$. Also, recall that a 
$2$-factor of $H$ is a collection of vertex-disjoint cycles from $H$ that 
spans all of its vertices.

\begin{theorem}
\label{decomp}
$G$ decomposes 
into $\floor{k/2}$ $2$-factors and, if $k$ is odd, a perfect matching as well. 
For $1\leq i\leq \floor{k/2}$, the edges of the $i$th $2$-factor are labeled 
with $\gamma_i$, and the edges of the perfect matching with $n/2$.
\end{theorem}
\begin{proof}
The $k$ edges incident to each vertex are labeled with distinct members of 
$\Gamma$ on their near ends. So each of the $\floor{k/2}$ label pairs asserted 
initially in Lemma~\ref{labeling} can be used to identify a different 
$2$-factor. Such $2$-factors encompass all of $G$, with the exception of the edges 
whose label pair is $\{n/2,n/2\}$ in the odd-$k$ case (again, in Lemma~\ref{labeling}).
But these clearly constitute a perfect matching in $G$.
\end{proof}

\section{The Necessity of a Hamiltonian Cycle}

In this section we assume that $G$ is connected and begin by asserting
a relationship between two vertices that belong to a same 
cycle of one of the $2$-factors established in Theorem~\ref{decomp}. Let us 
denote by $r(u)$ the rank of vertex $u$ in the cyclic ordering that underlies 
the CSD.

\begin{lemma}
\label{samecycle}
For $1\leq i\leq \floor{k/2}$, two vertices $u$ and $v$ belong to a common cycle of
the $2$-factor whose edges are labeled with $\gamma_i$ if and only if $r(v)
=r(u)\pn t{\gamma_i}$ for some integer $t\geq 0$. 
\end{lemma}
\begin{proof}
If $u$ and $v$ share a common $\gamma_i$-labeled cycle, then traversing the 
cycle from $u$ to $v$ in the direction that exits a vertex along the 
$\gamma_i$-labeled end of the edge adds (modulo $n$) $\gamma_i$ rank units to 
$r(u)$ for each edge traversed. Then there exists a nonnegative integer $t$ such 
that $r(v)=r(u)\pn t\gamma_i$.

Conversely, if $r(v)=r(u)\pn t\gamma_i$ for some $t\geq 0$, then,
since every vertex has an incident 
edge labeled with $\gamma_i$ on the near end, $v$ can be reached by a path that begins at $u$, 
exclusively uses edges labeled with $\gamma_i$, and has $t$ edges. Clearly, such a path
is part of a cycle of a $2$-factor whose edges are labeled with $\gamma_i$.   
\end{proof}

Recall now that two integers $a$ and $b$ are \emph{relative primes}, denoted by 
$a \perp b$, if $\gcd(a,b)=1$.

\begin{fact}
\label{numtheory}
Let $a$ and $b$ be integers. Then $a \perp n$ if and only if the smallest $b>0$ 
that satisfies $b \tn a=0$ is $b=n$.
\end{fact}

We are now in position to demonstrate that $G$ is Hamiltonian. We do this by 
splitting the proof into cases that bear on the relative primality between 
each of $\gamma_1,\ldots,\gamma_{\floor{k/2}}$ and $n$.

\begin{theorem}
\label{prime_gamma}
If there exists $\gamma_i \in \{\gamma_1,\ldots,\gamma_{\floor{k/2}}\}$ such 
that $\gamma_i \perp n$, then $G$ has a Hamiltonian cycle whose edges are 
labeled with $\gamma_i$.
\end{theorem}
\begin{proof}
By Fact~\ref{numtheory}, the smallest integer $t>0$ that satisfies 
$t\tn \gamma_i=0$ is $t=n$. In the same way, for any vertex $u$, the smallest 
$t>0$ that satisfies $r(u)\pn t\gamma_i=r(u)$ is also $t=n$. By Lemma~\ref{samecycle},
vertex $u$ is on an $n$-vertex cycle whose edges are labeled with $\gamma_i$.
\end{proof}

Before we proceed with the case in which no $\gamma_i \in \{\gamma_1,\ldots,
\gamma_{\floor{k/2}}\}$ is such that $\gamma_i\perp n$, we give a necessary 
condition for such a scenario to happen.\footnote{The
reader should note that Theorem~\ref{prime_gamma} and Lemma~\ref{cycles}
could be coalesced into one single result stating that the number of cycles in the
$2$-factor whose edges are labeled with $\gamma_i$ for $1\leq i\leq\floor{k/2}$ is
$\gcd(\gamma_i,n)$. We choose to do otherwise for clarity's sake only.}
We also prove a general property on the relative primality 
between the greatest common divisor of all the labels that are no greater than
$n/2$ (including, if applicable, $n/2$ itself) and $n$.

\begin{lemma}
\label{cycles}
If no $\gamma_i \in \{\gamma_1,\ldots,\gamma_{\floor{k/2}}\}$ exists such that
$\gamma_i \perp n$, then for $1\leq i\leq\floor{k/2}$ the edges of $G$ labeled with
$\gamma_i$ form a 
$2$-factor with $d_i$ cycles of the same length, where $d_i=\gcd(\gamma_i,n)>1$.
\end{lemma}
\begin{proof}
First rewrite $t\tn \gamma_i=0$ as $t\tnd \gamma_i /d_i=0$ and notice that 
$\gamma_i /d_i \perp n/d_i$. By Fact~\ref{numtheory}, it follows that $t=n/d_i$ is the 
smallest integer that satisfies $t\tnd \gamma_i /d_i=0$, and hence also 
$t\tn \gamma_i=0$. By Lemma~\ref{samecycle}, each cycle of the $2$-factor whose 
edges are labeled with $\gamma_i$ comprises $n/d_i$ vertices, so by Theorem~\ref{decomp}
the number of such cycles is $d_i$.
\end{proof}

\begin{lemma}
\label{gcd}
Let $d=\gcd(\gamma_1,\ldots,\gamma_{\ceil{k/2}})$. Then $d\perp n$.
\end{lemma}
\begin{proof}
Let $t_i\geq 0$, for $1\leq i\leq \ceil{k/2}$, be an integer. Thus, 
$\sum_i t_i\gamma_i$ is a multiple of $d$ and represents, for an arbitrary walk in 
$G$, the rank difference along the CSD's cyclic vertex ordering from the walk's 
initial vertex to its final vertex. To see why any walk is thus contemplated, 
notice that a walk that uses $t_i$ edges labeled with $n-\gamma_i$ on their
near ends as vertices are exited along the walk can be 
substituted for by another one that connects the same two vertices and uses 
$n/\gcd(\gamma_i,n)-t_i$ edges labeled with $\gamma_i$ instead. Therefore, for 
a walk between arbitrary vertices $u$ and $v$, the rank difference between these 
vertices in the cyclic ordering is given by $t\tn d$ for some
$t\in\{0,\ldots,n-1\}$. And, since $G$ is connected, every possible value of
$t\tn d$ (i.e., $0,\ldots,n-1$) must result from a distinct value of $t$.
If such is the case, then 
$\gcd(d,n)=1$, that is, $d \perp n$.\footnote{A formal proof of this implication 
can be found in Section~4.8 of \cite{knuth}.}
\end{proof}

It is important to recall that, when $k$ is odd, $\gamma_{\ceil{k/2}}=n/2$ by 
Lemma~\ref{labeling}. In this case, by Lemma~\ref{gcd} we must have $d=1$ (since 
$d$ divides $n/2$, therefore $n$ as well, which for $d>1$ contradicts the lemma).
The existence of a Hamiltonian cycle when none of $\gamma_1,\ldots,
\gamma_{\floor{k/2}}$ is relatively prime to $n$ can now be proven.

\begin{figure}[t]
\centering
\includegraphics[scale=0.75]{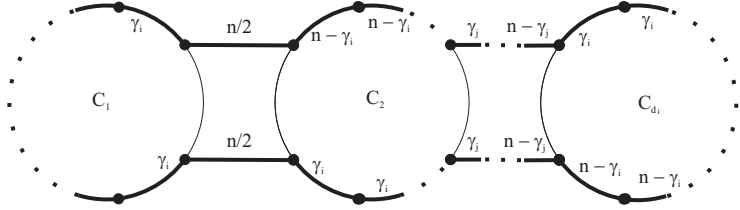}
\caption{Hamiltonian cycle using the cycles whose edges are labeled with 
$\gamma_i$ and edges labeled with $n/2$ and $\gamma_j$.}
\label{fig_hamilt2}
\end{figure}

\begin{theorem}
\label{not_prime_gamma}
If no $\gamma_i \in \{\gamma_1,\ldots,\gamma_{\floor{k/2}}\}$ exists such that 
$\gamma_i \perp n$, then $G$ has a Hamiltonian cycle.
\end{theorem}
\begin{proof}
Note, first, that every one of $\gamma_1,\ldots,\gamma_{\floor{k/2}}$ is
necessarily greater than $1$.
Let $d'=\gcd(\gamma_1,\ldots,\gamma_{\floor{k/2}})$. The proof is divided into 
two cases: $d'=1$ and $d'>1$.

Let $d'=1$. Then there exist $\gamma_i$ and $\gamma_j$ in $\{\gamma_1,\ldots,
\gamma_{\floor{k/2}}\}$ such that $\gamma_i \perp \gamma_j$. Suppose that two 
vertices $u$ and $v$ are connected by an edge labeled with $\gamma_i$ and belong 
to the same cycle of the $2$-factor whose edges are labeled with $\gamma_j$. In 
this case, we have $\gamma_i=t \tn \gamma_j$ for some integer $t\geq 0$, which 
is a contradiction, since $\gamma_i \perp \gamma_j$. It follows that the end 
vertices of an edge labeled with $\gamma_i$ belong to distinct cycles of the 
$2$-factor whose edges are labeled with $\gamma_j$. We can also say that, if $u$ 
and $v$ are adjacent on a cycle $C_1$ of the $2$-factor whose edges are labeled 
with $\gamma_i$, then there exist edges labeled with $\gamma_j$ that connect $u$ 
and $v$ to vertices $u'$ and $v'$, respectively, where $u'$ and $v'$ are 
adjacent on a cycle $C_2$ of the same $2$-factor. This is summarized in the 
equality
\begin{equation}
\label{equality}
\gamma_j\pn\gamma_i\pn(n-\gamma_j) = \gamma_i,
\end{equation}
which refers to the illustration in Figure~\ref{fig_path}.
Let then $C_1,\ldots,C_{d_i}$ be the cycles of the $2$-factor whose edges are labeled 
with $\gamma_i$ (by Lemma~\ref{cycles}, $d_i=\gcd(\gamma_i,n)$). We can easily 
identify a Hamiltonian cycle using $C_1,\ldots,C_{d_i}$
and interconnecting these cycles through the edges labeled with 
$\gamma_j$, as in Figure~\ref{fig_hamilt1}.

\begin{figure}[t]
\centering
\includegraphics[scale=0.75]{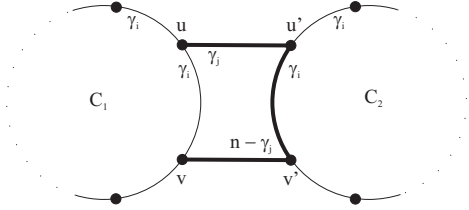}
\caption{A path connecting vertices $u$ and $v$ and whose edges are labeled 
with $\gamma_j$, $\gamma_i$, and $n-\gamma_j$ (in this order). Cycles $C_1$ and 
$C_2$ belong to the $2$-factor whose edges are labeled with $\gamma_i$.}
\label{fig_path}
\end{figure}

\begin{figure}[t]
\centering
\includegraphics[scale=0.75]{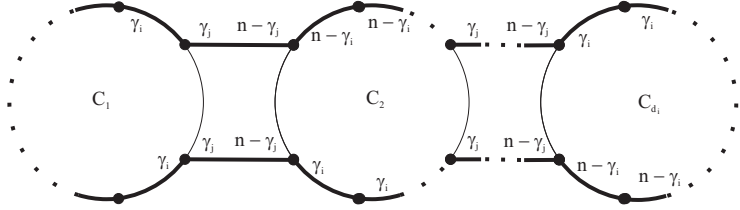}
\caption{Hamiltonian cycle using the cycles whose edges are labeled with 
$\gamma_i$ and edges labeled with $\gamma_j$.}
\label{fig_hamilt1}
\end{figure}

Now let $d'>1$. If no edges are labeled with $n/2$, then every path in $G$, say
from $u$ to $v$,
is such that $r(v)\mn r(u)$ is a multiple (modulo $n$) of $d'$. And since $G$ is
connected, this has to hold for all vertex pairs in the graph, even those whose
rank differences are not a multiple of $d'$. This is clearly contradictory, so
there have to exist
edges labeled with $n/2$ (in which case $k$ must be odd, by Lemma~\ref{labeling})
and we must have
$d'\perp n/2$ (by Lemma~\ref{gcd}, according to which $\gcd(d',n/2)\perp n$). 
This latter conclusion allows us to
substitute $n/2$ for $\gamma_j$ in (\ref{equality}), and then we see that the edges 
labeled with $n/2$ connect two distinct cycles of the $2$-factor whose edges are 
labeled with $\gamma_i$. If we take all such cycles and alternately interconnect 
them by an edge labeled with $n/2$ and another labeled with an 
appropriate $\gamma_j$,\footnote{It suffices for $\gamma_i$ and $\gamma_j$ not to
be multiples of each other. The case in which every one of $\gamma_1,\ldots,
\gamma_{\floor{k/2}}$ is a multiple of $\gamma_i$ has a trivial Hamiltonian cycle 
that uses only edges labeled with $\gamma_i$ and $n/2$.} then a 
Hamiltonian cycle is easily constructed, as in Figure~\ref{fig_hamilt2}.
\end{proof}

The following is then straightforward.

\begin{corollary}
\label{hamilt}
$G$ has a Hamiltonian cycle.
\end{corollary}
\begin{proof}
By Theorems~\ref{prime_gamma} and \ref{not_prime_gamma}.
\end{proof}

Note, finally, that even though the presence of a
Hamiltonian cycle is a necessary condition for an 
edge labeling to be an MCSD in connected regular graphs,
it is not a sufficient condition. 
In fact, it is easy to find $k$-regular graphs that have a Hamiltonian cycle but 
do not admit a CSD with $k$ labels, as the example in Figure~\ref{fig_example}(b).

\section{Deciding Whether an MCSD Exists}

Given an arbitrary set $\{\gamma_1,\ldots,\gamma_{\ceil{k/2}}\}$ of labels such
that $\gamma_1<\cdots<\gamma_{\ceil{k/2}}$, if none of its members is
greater than $n/2$ with $\gamma_{\ceil{k/2}}=n/2$ in the odd-$k$ case,
then one can easily (polynomially) generate a $k$-regular graph $H$ with an MCSD by simply 
arranging the $n$ vertices in a cyclic ordering and, for each $\gamma_i \in 
\{\gamma_1,\ldots,\gamma_{\ceil{k/2}}\}$, connecting pairs of vertices whose 
ranks in the ordering differ by $\gamma_i$ and labeling the resulting edges 
with $\{\gamma_i,n-\gamma_i\}$ appropriately.
In order for $H$ to be connected, by Lemma~\ref{gcd} we require in addition that
$\gcd(\gamma_1,\ldots,\gamma_{\ceil{k/2}})\perp n$.
Thus, a possible
direction towards the development of an algorithm to check whether a given
$k$-regular graph $G$ admits an MCSD is to generate $H$ in this way for 
every pertinent set of labels,\footnote{It is easy to see that $H$ is unique for
a given set of labels.} and then to check whether $H$ is 
isomorphic to $G$.

When we fix the input graph's degree (i.e., $k$ is a constant), the 
maximum number of candidate labelings to be checked if we ignore the restriction that
$\gcd(\gamma_1,\ldots,\gamma_{\ceil{k/2}})\perp n$ in the connected case is
\[ {\floor{n/2}\choose \floor{k/2}}=O(n^k), \]
a polynomial in $n$. In \cite{luks}, the isomorphism of graphs of bounded degree 
was shown to be testable in polynomial time. Thus, we can decide whether $G$ 
admits an MCSD also polynomially.

\begin{figure}[t]
\centering
\begin{tabular}{c@{\hspace{1.5cm}}c}
\includegraphics[scale=0.75]{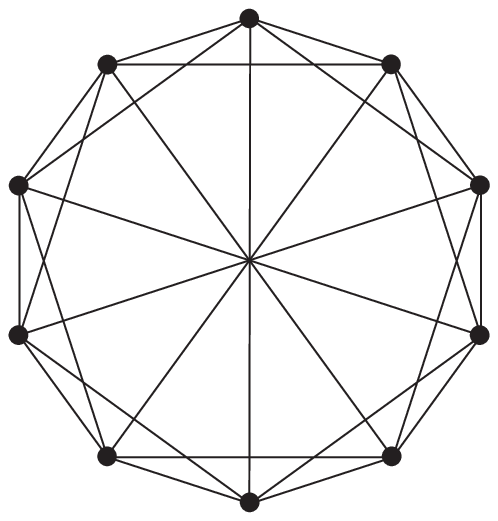}&
\includegraphics[scale=0.75]{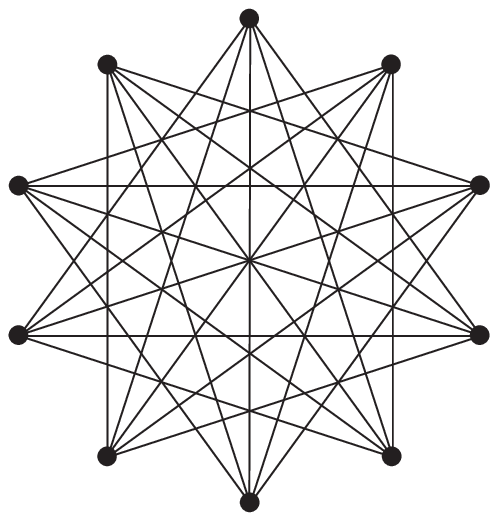}\\
{\small (a)}&{\small (b)}
\end{tabular}
\caption{Isomorphic graphs with equivalent MCSD labelings, based on 
the label sets $\{1,2,5\}$ (a) and $\{3,4,5\}$ (b).}
\label{fig_equiv}
\end{figure}

The complexity of the overall MCSD test can be clearly improved if we consider 
the possible 
isomorphism between graphs generated from distinct valid label sets. We say 
that two distinct labelings $\lambda$ and $\lambda'$ are \emph{equivalent}, 
denoted by $\lambda \equiv \lambda'$, if they generate isomorphic graphs. For 
example, it can be easily seen that the labelings $\lambda$ and $\lambda'$, 
drawing respectively on the label sets $\{1,2,5\}$ and $\{3,4,5\}$, 
generate isomorphic $5$-regular graphs on $10$ vertices (see 
Figure~\ref{fig_equiv}), so $\lambda \equiv \lambda'$. 
Let us then consider a transformation of $\lambda$ into $\lambda'$ that 
preserves the MCSD property. In what follows, $\lambda$ draws on the 
set $\{\gamma_1,\ldots,\gamma_{\ceil{k/2}}\}$ for labels, $\lambda'$ on 
$\{\gamma_1',\ldots,\gamma_{\ceil{k/2}}'\}$.

\begin{theorem}
\label{transform}
Let $\alpha<n/2$ be an integer such that $\alpha \perp n$. For $1\leq i\leq 
\ceil{k/2}$, let
\[ \gamma_i'= \left\{\begin{array}{ll}
\alpha\tn \gamma_i & \mbox{\rm if $\alpha\tn \gamma_i\leq n/2$} \\
n-\alpha\tn \gamma_i & \mbox{\rm if $\alpha\tn \gamma_i>n/2$}. \\
\end{array} \right. \]
If $\lambda$ is an MCSD for $G$, then so is $\lambda'$.
\end{theorem} 
\begin{proof}
It suffices that we argue that no member of $\{\gamma_1',\ldots,\gamma_{\ceil{k/2}}'\}$
is greater than $n/2$ with $\gamma_{\ceil{k/2}}=n/2$ for $k$ odd,
that every two members of this set are distinct, and also that
the vertices can be rearranged cyclically so that $\lambda'$ is indeed a CSD. The first of
these properties holds trivially and the second follows from well-known number-theoretic
properties.\footnote{We once again refer the reader to Section~4.8 of \cite{knuth}.}
As for the third property, clearly it suffices for the vertices to be arranged into a
cyclic ordering in which vertex $u$ has rank $r'(u)$ such that
$r'(u)=\alpha\tn r(u)$.
\end{proof}  

It is easy to see that any $\alpha>n/2$ would produce the same results as its 
symmetric modulo $n$. We can also see that any $\lambda$ comprising a $\gamma_i$ 
such that $\gamma_i \perp n$ can yield a $\lambda'$ with $\gamma_i'=1$. For 
such, it is sufficient to take $\alpha$ as the  multiplicative inverse (modulo 
$n$) of $\gamma_i$.   

It is also curious to note that the transformation in Theorem~\ref{transform}
ensures that $\gamma_i\perp n$ if and only if $\gamma_i'\perp n$. To see this,
consider for example the case of $\gamma_i'=\alpha\tn\gamma_i$. By using
Euclid's Theorem \cite{knuth} and the fact that $\alpha\perp n$ in succession,
we have $\gcd(\alpha\tn\gamma_i,n)=\gcd(\alpha\gamma_i,n)=\gcd(\gamma_i,n)$,
thence $\gcd(\gamma_i',n)=\gcd(\gamma_i,n)$.

\section{MCSD's and Circulant Graphs}
\label{circ}

There is a clear equivalence between circulant graphs and
regular graphs that admit an MCSD. We describe it formally in the following 
theorem.

\begin{theorem}
\label{equiv}
$G$ is circulant of generator set $S$ if and only if it is $|S|$-regular and
admits an MCSD.
\end{theorem}
\begin{proof}
Let $G$ be a circulant graph of generator set $S$. Then $uv\in E(G)$ if and only
if there exists $s\in S$ such that $v=u\pn s$.
Let $\lambda$ be a labeling for $G$ such that 
$\lambda_u(uv)=s$. Since the vertices of $G$ are elements of $\mathbb{Z}_n$, 
they already have a natural cyclic ordering in which $r(u)=u$ for all $u \in 
V(G)$. So $\lambda_u(uv)=r(v) \mn r(u)$ and $\lambda$ is a CSD of $G$. Also, as 
every vertex is connected to the vertex ranking $s$ higher (modulo $n$) than
itself for every $s\in S$, $G$ is $|S|$-regular and $\lambda$ uses $|S|$ labels
(thence the CSD is minimal).

Conversely, if $G$ is a $k$-regular graph that admits an MCSD, then
$|\Gamma|=k$ and there exists 
a cyclic ordering of the vertices such that each edge $uv$ is labeled with 
$\lambda_u(uv)=r(v) \mn r(u)$, where $0\leq r(u) \leq n-1$. Letting
$V(G)=\mathbb{Z}_n$ so that $u=r(u)$ and $S=\Gamma$ yields $\lambda_u(uv)=v\mn u$
for all $uv\in E(G)$, thence $v=u\pn\lambda_u(uv)$. $G$ is therefore circulant
of generator set $S$.
\end{proof}

One first example of how Theorem~\ref{equiv} sheds new light on the two concepts
involved comes from considering the result on circulant graphs in 
\cite{boesch}, which implies in the connected case that $\gcd(s_0,\ldots,s_k,n)=1$, where 
$s_i$, for $0\leq i \leq k$, is an element of the set of generators.
This is of course coherent with
Lemma~\ref{gcd} and Corollary~\ref{hamilt}, and a straightforward 
implication of the general result of \cite{boesch}
is that a circulant graph is Hamiltonian if and only if it 
is connected \cite{burkard}. Our approach introduces new ways of 
constructing Hamiltonian cycles in this case.

Another interesting insight is the following. An $n \times n$ matrix
is said to be \emph{circulant} if its $i$th line is the 
cyclic shift of the first line by $i$ positions,
where $0\leq i\leq n-1$. Another characterization of 
circulant graphs is that a graph is circulant if its 
adjacency matrix is circulant. By 
the equivalence established in Theorem~\ref{equiv}, it then becomes possible to
approach the problem of recognizing regular graphs that admit an MCSD along
a different route: since it is well-known that the 
isomorphism between two graphs $G$ and $H$ can be viewed as a permutation of 
lines and columns of the adjacency matrix of $G$ ($A(G)$) that generates that of
$H$, we can test whether a regular graph $G$ admits an MCSD by
finding a permutation of lines and columns of $A(G)$ such that the resulting 
matrix is circulant.

We note, in addition, that the transformation defined in Theorem~\ref{transform}
also has an analogue in the 
literature on circulant graphs. Let $G$ and $H$ be circulant graphs such that their 
sets of generators are $R$ and $S$, respectively.
We say that $R$ and $S$ are $proportional$, denoted 
by $R \sim S$, if, for some $a\perp n$, $r=a\tn s$ bijectively for $r\in R$ and $s\in S$.
Clearly, if $R\sim S$ then 
$G$ is isomorphic to $H$.
The converse statement was conjectured in \cite{adam} and is known as 
\emph{\'Ad\'am's conjecture}. However, in \cite{elspas} the conjecture was 
proven false.

The problem of recognizing circulant graphs, finally, is probably the one most affected 
by the equivalence of Theorem~\ref{equiv}. Even though the algorithm we suggest 
to test whether a $k$-regular graph admits an MCSD is polynomial only for
fixed $k$, when applied to the context of circulant graphs it is the only known 
result on arbitrary topologies (without any restrictions on the structure or the 
number of vertices) for that class. 

\subsection*{Acknowledgments}

The authors acknowledge partial support from CNPq, CAPES, and a FAPERJ BBP
grant.

\bibliography{wg2005}
\bibliographystyle{plain}

\end{document}